\documentclass[12pt,preprint]{aastex}
\newcommand{\beq}{\begin{equation}}
\newcommand{\beqa}{\begin{eqnarray}}
\newcommand{\eeq}{\end{equation}}
\newcommand{\eeqa}{\end{eqnarray}}
\newcommand{\simg}{\gtrsim}
\newcommand{\siml}{\lesssim}
\newcommand{\meszaros}{M${\acute {\rm e}}$sz${\acute {\rm a}}$ros}
\shorttitle{Spatial Correlation of MBH Mergers}
\shortauthors{Ioka \& \meszaros}
\begin{document}
\title{
Spatial Correlation of Massive Black Hole Mergers:
Probing the Formation Mechanism and Reionization
}
\author{
Kunihito Ioka\footnote{
Physics Department and Center for Gravitational Wave Physics, 
104 Davey Laboratory, Pennsylvania State University, University Park,
PA 16802}
and Peter \meszaros$^{1,}$\footnote{
Department of Astronomy and Astrophysics, 525 Davey Laboratory,
Pennsylvania State University, University Park, PA 16802}
}

\begin{abstract}
We consider the spatial clustering of massive black hole (MBH) mergers, and
discuss possible ways to use gravitational wave observations in the LISA 
and DECIGO/BBO range for obtaining cosmological and cosmogonical information. 
Constraints on large scale structure (LSS) and merger histories may be 
possible through the detection of an alignment of the GW polarization 
direction with principal axes of the LSS. Constraints on the merger physics 
and the reionization epoch may be obtained by GW measurements of MBH 
correlation lengths, in the case when the MBH angular momentum loss occurs 
through gas drag. Such measurements would provide information about the 
LSS and the reionization epoch, as well as about the astrophysics of MBH 
mergers, additional to and independent of that obtained from 
electromagnetic signals.
\end{abstract}


\section{Introduction and summary}

Gravitational waves (GWs) will provide new eyes for studying the universe in 
the 21st century. Ground-based laser interferometers such as TAMA300 and LIGO 
have already begun operations in the  $10$ Hz -- kHz band, while the Laser 
Interferometer Space Antenna (LISA)\footnote{http://lisa.jpl.nasa.gov/index.html}, 
covering the $10^{-4}$ -- $10^{-2}$ Hz band, will be launched in $2011$. Other
recently proposed facilities are the Decihertz Interferometer Gravitational Wave 
Observatory (DECIGO) \citep{seto01} and the Big Bang Observer (BBO)
\footnote{http://universe.gsfc.nasa.gov/}, operating around decihertz 
frequencies ($10^{-2}$ -- $10$ Hz). These detectors may determine the spatial 
distribution of GW sources, which should contain important cosmological as 
well as astrophysical information \citep[e.g.,][]{ioka00,seto05}.

In this paper we propose novel ways to use a 3D map of GW sources, in particular
of Massive Black Hole (MBH) mergers. MBH mergers are one of the most powerful 
GW sources in the universe. LISA may detect MBH mergers involving masses 
$10^{3} \sim 10^{7} M_{\odot}$ out to high redshifts, while DECIGO/BBO may 
localize $1 \sim 10^{3} M_{\odot}$ MBH mergers with an arcmin resolution
\citep{takahashi03}. They can also measure distances to MBH binaries through
the change of the GW frequency during the observation.

The MBH merger history has large uncertainties derived from the seed MBH formation 
and the angular momentum evolution of MBH binaries. The presence of high redshift 
quasars may imply that seed MBHs were formed much before the cosmological 
reionization epoch as Population III (Pop.III) remnants 
\citep{madau01,volonteri03}. In such a scenario the event rate could reach 
$\sim 10^{4}$ events per yr \citep[\S~\ref{sec:event};][]{islam04}, which 
makes them potentially interesting for doing cosmology. Although we will find 
it impossible to measure the spatial correlation of host galaxies of MBH 
mergers through GWs alone (\S~\ref{sec:event}), we may still use the 3D maps 
of MBH mergers to study the MBH evolution.

One possible way to extract information from a 3D map of GW sources is the 
cross-correlation of GWs with electromagnetic (EM) sources (\S~\ref{sec:align}). 
We will suggest that the orbital axis of the MBH binary can be aligned with a
principal axis of the large scale structure (LSS) in which the binary resides, 
and that this may be detectable with LISA and DECIGO/BBO. Such a detection would 
provide useful clues to the angular momentum gains of MBH binaries.

GW sources could have a larger correlation length than that of host galaxies,
and in this case we may detect a spatial correlation of GW sources
(\S~\ref{sec:gas}). We will suggest that the MBH merger rate may drop inside 
an ionized bubble, since ionizing photons heat the gas which could be responsible 
for the angular momentum loss of MBH binaries. Since ionized bubbles may be 
larger than the LSS scales at the end of reionization, we could detect the spatial 
correlation of MBH mergers and thus probe the effectiveness of MBH binary angular 
momentum loss via gas. The merger rate may also drop after reionization, thus
providing an independent marker for this important epoch. 

\section{Event rate and correlation length}\label{sec:event}

We first estimate the event rate of MBH mergers, adopting throughout a 
$\Lambda$CDM cosmology with $(\Omega_{m}, \Omega_{\Lambda}, \Omega_{b}, h, 
\sigma_{8}) =(0.27, 0.73, 0.044, 0.71, 0.84)$ \citep{spergel03}. Since 
little is known about the MBH merger history, we concentrate on a model 
in which MBH mergers are associated with the mergers of host halos of dark 
matter \citep{begelman80,volonteri03}. By using the extended \citet{ps74} 
(PS) formalism, we can estimate the halo merger rate as in equation 
(\ref{eq:rate}) \citep{lacey94,haehnelt94}.

The comoving number density of halos in the mass range $M \sim M+dM$ at 
redshift $z$ is given by the PS mass function as
\beqa
\frac{dn_{\rm PS}}{dM}(M,z)dM
=\sqrt{\frac{2}{\pi}} \frac{\rho_{m}}{M}
\frac{\delta_{c}(z)}{\sigma^{2}}
\left|\frac{d\sigma}{dM}\right|
\exp \left[-\frac{\delta_{c}^{2}(z)}{2\sigma^{2}}\right] dM,
\eeqa
where $\rho_{m}=3 H_{0}^{2} \Omega_{m}/8\pi G \simeq 2.78 \times 10^{11} 
\Omega_{m} h^{2} M_{\odot}$ Mpc$^{-3}$ is the mean comoving matter density,
$\sigma \equiv \sigma(M)$ is the rms mass fluctuation on a mass scale $M$
at the present epoch, $\delta_{c}(z) \equiv \delta_{c} D(z=0)/D(z)$ is the 
critical linear overdensity evaluated at present for a spherical 
perturbation to collapse at $z$, and $D(z)$ is the linear growth rate.
These halos of mass $M$ are incorporated into a larger halo of mass
$M_{2}\sim M_{2}+dM_{2}$ at redshift $z$ at a rate given by
\beqa
\frac{d^{2} f}{dM_{2} dz}(M \to M_{2};z)dM_{2}
&=&\frac{1}{\sqrt{2\pi}}
\left[\frac{\sigma^{2}}{\sigma_{2}^{2}(\sigma^{2}-\sigma_{2}^{2})}\right]^{3/2}
\frac{d\delta_{c}(z)}{dz}
\left|\frac{d\sigma_{2}^{2}}{dM_{2}}\right|
\nonumber\\
&\times& \exp \left[-\frac{(\sigma^{2}-\sigma_{2}^{2})\delta_{c}^{2}(z)}
{2\sigma^{2}\sigma_{2}^{2}}\right] dM_{2},
\eeqa
where $\sigma_{2} \equiv \sigma(M_{2})$ \citep{lacey94}. Therefore the 
merger rate of halos of mass larger than $M_{\min}$ with halos of the same 
mass or larger per comoving volume is given by
\beqa
\frac{dn_{\rm merg}}{dt}(z)=
\int_{M_{\min}}^{\infty} dM
\frac{dn_{\rm PS}}{dM}(M,z)
\int_{2M}^{\infty} dM_{2}
\frac{d^{2} f}{dM_{2} dz}(M \to M_{2};z)
\left|\frac{dz}{dt}\right|.
\label{eq:dnmdt}
\eeqa
This yields the total number of merger events per observed time 
$t_{\rm obs}$ per redshift $z$ as
\beqa
\frac{dN_{\rm merg}}{dz dt_{\rm obs}}(z)
=\frac{1}{1+z}
\frac{dn_{\rm merg}}{dt}
\frac{dV_{c}}{dz},
\label{eq:rate}
\eeqa
where $dV_{c}/dz=4 \pi c d_{L}^{2} |dt/dz|/(1+z)$ is the comoving volume 
per redshift, $d_{L}=c (1+z) \int_{0}^{z} (1+z) |dt/dz| dz$ is the 
luminosity distance, and 
$|dt/dz|^{-1}=(1+z)H_{0}\sqrt{\Omega_{m}(1+z)^{3}+\Omega_{\Lambda}}$.
We follow \citet{ks96} to estimate $\delta_{c}$ and $\sigma(M)$.

We note that the merger rate in equation (\ref{eq:dnmdt}) is dominated 
by equal mass halos ($M_{2}\sim 2M$). Thus the rate is not reduced much 
even if dynamical friction is inefficient in initially bringing halos
of small mass ratios together \citep{lacey94,yu02}. The dynamical friction 
also becomes ineffective as the MBH binary hardens. It is unclear how the 
MBH binary loses its angular momentum afterward. The two main 
possibilities are through interactions with stars or with gas. Stars may 
not be effective since the MBH binary will quickly eject all stars on 
intersecting orbits \citep{milos03,merritt04}, while gas may reduce the 
binary separation efficiently \citep{gould00,armitage02,escala04}. Here we assume that 
two MBHs coalesce soon after the host halos merge. In \S~\ref{sec:gas} 
we will propose observational methods to test whether gas is responsible 
for the angular momentum losses.

The event rate in equation (\ref{eq:dnmdt}) is dominated by halos with 
the minimum mass $M \sim M_{\min}$. It is not well known what fraction 
of galaxies (and down to what mass) harbor MBHs, although almost all 
nearby massive galaxies have central MBHs \citep{magorrian98}. Here we 
assume that all halos of mass above $M_{\min}(z)$ have a central MBH and 
consider three possibilities for $M_{\min}(z)$: (a) The minimum halos 
have a virial temperature $T_{\rm vir} \sim 10^{4}$ K, above which baryons 
can cool via atomic hydrogen lines \citep{barkana01,menou01}. We 
calculate $T_{\rm vir}$ according to \citet{ks96}. (b) The minimum halos 
have $T_{\rm vir} \sim 10^{3}$ K, above which molecular hydrogen 
(H$_{2}$) cooling is possible \citep{barkana01,bromm04}. (c) The minimum 
halos contain more than one Pop.III MBH on average. We assume that one 
MBH forms in each $3 \sigma$ halo at $z=20$, and the mean number of MBHs 
in a given halo is calculated by equation (3) of \citet{madau01}.

\begin{figure}
\plotone{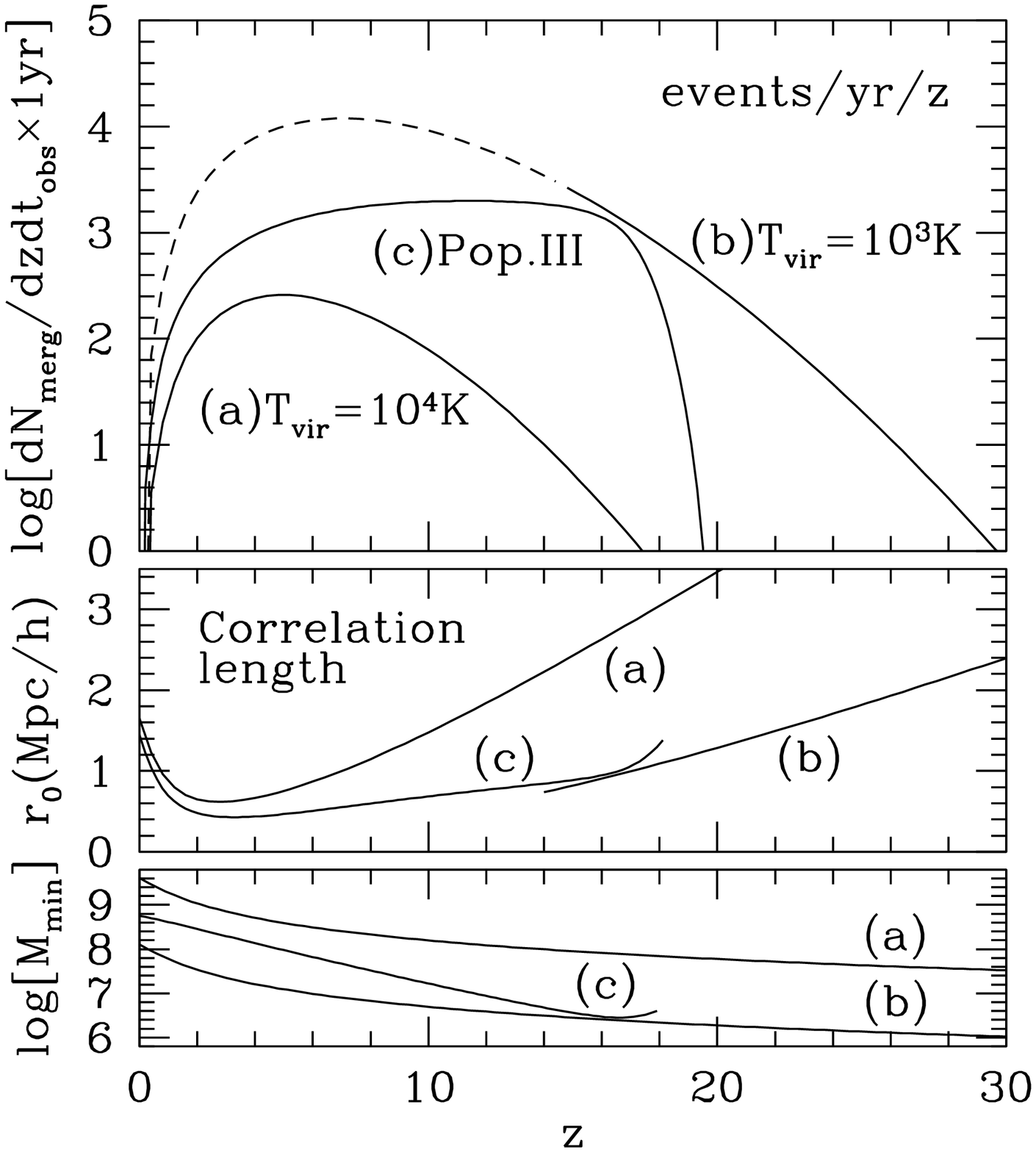}
\caption{\label{fig:rate}
Upper panel -- Number of MBH merger events per yr per unit redshift $z$ 
as a function of $z$. Three cases are considered for the minimum halos 
that harbor a central MBH: 
(a) The virial temperature $T_{\rm vir}$ of minimum halos is $10^{4}$ K,
above which atomic cooling is effective.
(b) The minimum halos have $T_{\rm vir}=10^{3}$ K,
above which molecular cooling is possible. This is shown with a dashed 
line for $z<15$, since molecular cooling may be ineffective in this range.
(c) The minimum halos contain more than one Population III MBH on average,
where we assume that one MBH forms in each $3 \sigma$ halo at $z=20$.
Middle panel -- Correlation length of the MBH mergers as a function of $z$.
Case (b) is plotted only for $z \simg 14$, and case (c) only for 
$z \siml 18$ (see text).
Lower panel -- Minimum halo mass that harbors a central MBH $M_{\rm
 min}(z)$ as a function of $z$.
}
\end{figure}

Figure~\ref{fig:rate} shows the number of merger events per yr per unit $z$
as a function of $z$. We can see that in total $\sim 10^{3}$ events/yr are 
expected in case (a), and $\sim 10^{4}$ events/yr in the case (c). In the 
case (b) we plot only the region of $z \simg 15$ with a solid line because
H$_{2}$ is fragile and can be destroyed by photons in the Lyman-Werner
(11.2-13.6 eV) bands well before the cosmological reionization
\citep{haiman00,omukai99}. The H$_{2}$ history is very uncertain because
free electrons can promote H$_{2}$ formation \citep{bromm04}, so that
$\simg 10^{3}$ events/yr at $z \simg 15$ is not implausible. 
The mass density of MBHs in the case (b) exceeds that found in the
nuclei of nearby galaxies \citep{madau01,yu02b}. Thus in the case (b) the
merging efficiency should drop at low redshift and many MBHs are to be
wandering without being observed.
The event rate in the case (c) is larger 
than that in the case (a). This suggests that the MBH merger might happen 
in nonluminous dwarf galaxies, which have not been explored yet.

The MBH mass, especially in small halos, is quite uncertain. If we 
extrapolate the relation of MBH and halo mass \citep{ferrarese02}, we 
have $M_{\rm MBH} \sim 10^{7} M_{\odot} 
({T_{\rm vir}}/{10^{6}{\rm \ K}})^{\alpha}$ where $\alpha$ ranges from 
$\sim 5/2$ \citep{merritt01} to $\sim 2$ \citep{gebhardt00}. For 
$\alpha \sim 2$ the MBH mass in halos of $T_{\rm vir}\sim 10^{4}$K is 
$\sim 10^{3} M_{\odot}$, so that almost all events of case (a) may be 
detectable by LISA \citep[e.g.,][]{hughes02}. For $\alpha \sim 5/2$, 
we have $M_{\rm MBH} \sim 10^{2} M_{\odot}$. This is below the sensitivity 
of LISA but above that of DECIGO/BBO \citep{takahashi03}. In the case (c), 
the Pop.III MBH mass is also unclear. Depending on the gas accretion onto 
a protostellar core, the final mass of a Pop.III star and its end product 
may reach $\sim 10^{3} M_{\odot}$ \citep{omukai03}. In this case LISA may 
detect almost all Pop.III MBH mergers.

The first question to be addressed is whether the correlation length of 
the MBH mergers is measurable or not. Since the merger rate is dominated 
by the minimum halos of mass $M_{\min}$, the correlation length $r_{0}$ 
of the MBH mergers is roughly that of the minimum halos. We can then 
estimate the correlation length $r_{0}$ by\footnote{
The length $r_{0}$ is not exactly the same as the correlation length $r_{1}$
at which the correlation function is unity $\xi(r_{1})=1$. 
However the difference is negligible for our discussions.}
\beqa
b(M_{\min},z) \sigma(r_{0}) D(z)/D(z=0)=1,
\eeqa
where $b(M,z)=1+{\delta_{c}}^{-1}\left[{\delta_{c}^{2}(z)}{\sigma^{-2}}-1 \right]$
is the bias parameter for halos of mass $M$ at redshift $z$ \citep{mo96}.
Figure~\ref{fig:rate} shows the correlation length $r_{0}$ as a function of $z$.
Whether the correlation length $r_{0}$ is measurable or not is determined 
by the number $N_{\rm pair}$ of merger pairs whose distances are less than $r_{0}$.
We need at least $N_{\rm pair} \simg 10$ for a 3$\sigma$ detection since 
the Poisson error of the correlation function is given by 
$\Delta \xi/\xi \sim N_{\rm pair}^{-1/2}$ \citep[][\S~48]{peebles80}.
Unfortunately we can show that $N_{\rm pair}$ at each redshift bin 
$\delta z\sim 1$ is less than unity even if we observe for 10 yrs 
(even for the case (b)). Therefore we conclude that it is impossible to 
measure the correlation length $r_{0}$ of the host halos, using the MBH 
merger GWs, regardless of the detector sensitivity.

However, there are other types of spatial information and other means to
extract them involving MBH merger GWs. One of these is the exploitation of 
cross-correlations between GWs with EM sources (\S~\ref{sec:align}). 
Another is in the case when the spatial distribution of MBH mergers has 
a larger characteristic scale than the correlation length $r_{0}$ of the 
associated halos (\S~\ref{sec:gas}, see also \cite{khlopov05}). 
We discuss these two cases below.

\section{Orbital alignment with large scale structures}\label{sec:align}

\begin{figure}
\plotone{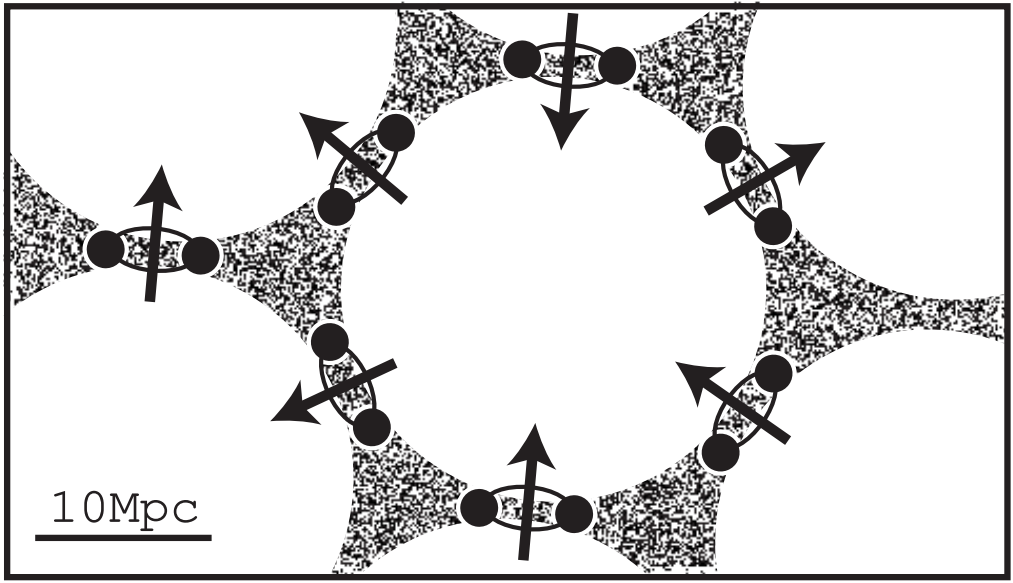}
\caption{\label{fig:align}
Schematic figure of LSSs (hatched region) and MBH binaries, illustrating
the preference for MBH binary axes to be normal to the filaments/sheets 
of LSSs. The sizes of MBH binaries are magnified to make them clear.
}
\end{figure}

Among various ways to cross-correlate GWs with EM sources, the one
involving the GW polarization and the EM LSS may be interesting. In the 
hierarchical structure formation scenario, sheet- or filament-like LSSs are 
initially formed. Such LSSs have a preferred direction, which could imprint a 
preferred orientation of the MBH binary relative to the LSS (see 
Figure~\ref{fig:align}). The orientation 
of the MBH binary may be determined through the GW polarization \citep{cutler98},
while the associated LSS can be mapped with wide-field cameras such as Suprime-Cam 
on the Subaru Telescope \citep{kodama01,ouchi05}. This offers the prospect of
correlating the MBH binary axis with the direction of the LSS.

The number of events needed to detect a correlation may be small. Let 
$\theta_{i}$ $(0^{\circ} \le \theta_{i} < 90^{\circ}, i=1, \cdots, N)$ be 
an angle between the binary axis and the associated filament direction.
(We focus on filaments to be concrete.) The anisotropy may then be 
quantified by \citep{struble85}
\beqa
\delta=\sum_{i}\frac{\theta_{i}}{N}-45.
\eeqa
For an isotropic distribution, $\delta \sim 0$ and the standard deviation 
would be $\sigma_{\theta}=90/(12 N)^{1/2}$ deg. Thus only $N \sim 10$ events 
may be enough for a 5$\sigma$ detection if the correlation is strong 
($\theta_{i} \sim 0^{\circ}$ or $90^{\circ}$).
However a $100\%$ alignment would be unlikely.
If the alignment is $\sim 20\%$ (which corresponds $\delta \sim 10$) as suggested 
for the alignment between galaxies and the LSS
\citep{pimbblet05,west95,lee02},
we need $N \sim 200$ events for a 5$\sigma$ detection.
Note that the alignment between the radio jets and
the galaxy disk has not been measured, but this is still consistent with
the $\sim 20\%$ alignment because of the Poisson error \citep{kinney00}.

We may apply the discussions on the alignment of galaxy spin axes with 
LSSs \citep[e.g.,][]{lee02,navarro04,west94}. In hierarchical models of 
galaxy formation, the galactic angular momentum is produced by tidal 
torques from the surrounding matter in the early protogalactic stage
\citep{peebles69,doro70,white84}. The tidal torque theory predicts that 
galaxy spins tend to align with the intermediate principal axis of the LSS,
suggesting that the MBH binary may also favor an orbital axis normal to 
the LSS filaments $\theta_{i} \sim 90^{\circ}$. Alternatively, the binary 
may have $\theta_{i} \sim 0^{\circ}$ if galaxies merge in the direction of 
the collapse from a sheet to a filament. Therefore the detection of such 
correlations would be useful for probing the MBH merger history.

A typical comoving scale of LSS at $z \sim 1$ is $\sim 20$ Mpc, which is 
about $\sim 1^{\circ}$ on the sky and $\Delta z \sim 0.01$ at $z\sim 1$.
Although LISA may not attain such precision in the localization of MBH 
mergers of mass $\sim 10^{3} M_{\odot}$ \citep{hughes02,cutler98,seto04},
we may be able to identify associated LSSs if we can find EM counterparts 
to the MBH merger \citep{tipler75,milos04}. On the other hand DECIGO/BBO 
should be able to identify associated LSSs because of its arcmin 
resolution \citep{takahashi03}. 
We may determine the orientation of the LSSs at the position of the MBH
merger with wide-field telescopes one by one, or we may have already
known the LSSs from future high-redshift surveys or 21 cm surveys.

\section{Effect of photoionization on MBH-drag}\label{sec:gas}

One of the major problems in the MBH evolution is uncertainty about 
how the MBH binary loses its angular momentum. The two main candidates 
for extracting the angular momentum are stars and gas. With stars there 
is the so called final parsec problem, in which the binary quickly ejects 
all stars on intersecting orbits and cuts off the supply of stars 
\citep{milos03,merritt04}, although this regions may be refilled via 
star-star encounters \citep{milos03} or chaotic orbits \citep{mpoon04}.
On the other hand, gas may be efficient in reducing the binary 
separation \citep{gould00,armitage02,escala04}.
Here we assume that gas is responsible for the angular momentum losses,
and show that this assumption may be verified or disproved by
the spatial distribution of MBH mergers.

We point out that a photoionizing background would heat the gas that 
is assumed to be responsible for the angular momentum losses and 
would inhibit the drag exerted on MBHs by gas, especially in small halos.
A photoionizing background photoevaporates gas in mini halos 
with virial temperatures $T_{\rm vir} \siml 10^{4}$ K
and substantially reduces the cool gas in halos with 
$10^{4} {\rm \ K} \siml T_{\rm vir} \siml 10^{5}$ K
\citep{ikeuchi86,efstathiou92,thoul96}.
Since the amount of MBH-dragging gas is reduced,
the MBH merger rate should drop in an ionized bubble,
possibly to almost zero because the event rate is dominated by small halos.
This would result in a spatial correlation of MBH mergers, and its detection 
would indicate that MBH binaries lose their angular momentum via gas.

The evolution of ionized bubbles during cosmological reionization
is still unclear, despite intensive studies \citep{barkana01}.
The analysis of Ly$\alpha$ spectra in the highest redshift quasars
indicates that the reionization ends at $z\sim 6$ \citep{fan02},
while the WMAP polarization data imply an onset of reionization at
a much higher redshift $z \sim 17 \pm 5$ \citep{kogut03}.  Also under 
discussion is whether the reionization proceeds from high- to low-density 
regions \citep{furlanetto04} or from low- to high-density regions \citep{miralda00}.

Recent observations and simulations suggest that the ionized bubble size 
at the end of their overlap is essentially determined by the finite bubble 
size and the cosmic variance \citep{wyithe04,furlanetto04}.
A comoving size $R$ of a finite bubble occupies a redshift interval of
\beqa
\Delta z=\frac{R}{c(1+z)}\left|\frac{dz}{dt}\right|,
\label{eq:finite}
\eeqa
which produces a scatter $\sim \Delta z$ of reionization redshift.
On the other hand, reionization in a certain region would be completed
when the fraction of collapsed mass exceeds a critical value in this region.
Because of the cosmic variance, a different region reaches the critical point at 
different redshift with a scatter
\beqa
\frac{\Delta z}{1+z}
=\frac{\bar \delta_{R}}{\delta_{c}(z)}-1+
\sqrt{1-\frac{\sigma^{2}_{R}}{\sigma^{2}_{R_{0}}}},
\label{eq:variance}
\eeqa
where $\sigma_{R} \equiv \sigma[M(R)]$ is 
the rms mass fluctuation over the comoving radius $R$ at the present epoch,
the mass $M(R_{0})$ within $R_{0}$ is the minimum galaxy mass
and $\bar \delta_{R}$ is the mean overdensity on the radius $R$.
Here we set $\bar \delta_{R}=\sigma_{R}$, and 
choose $M(R_{0})$ to have $T_{\rm vir}=10^{4}$ K.

\begin{figure}
\plotone{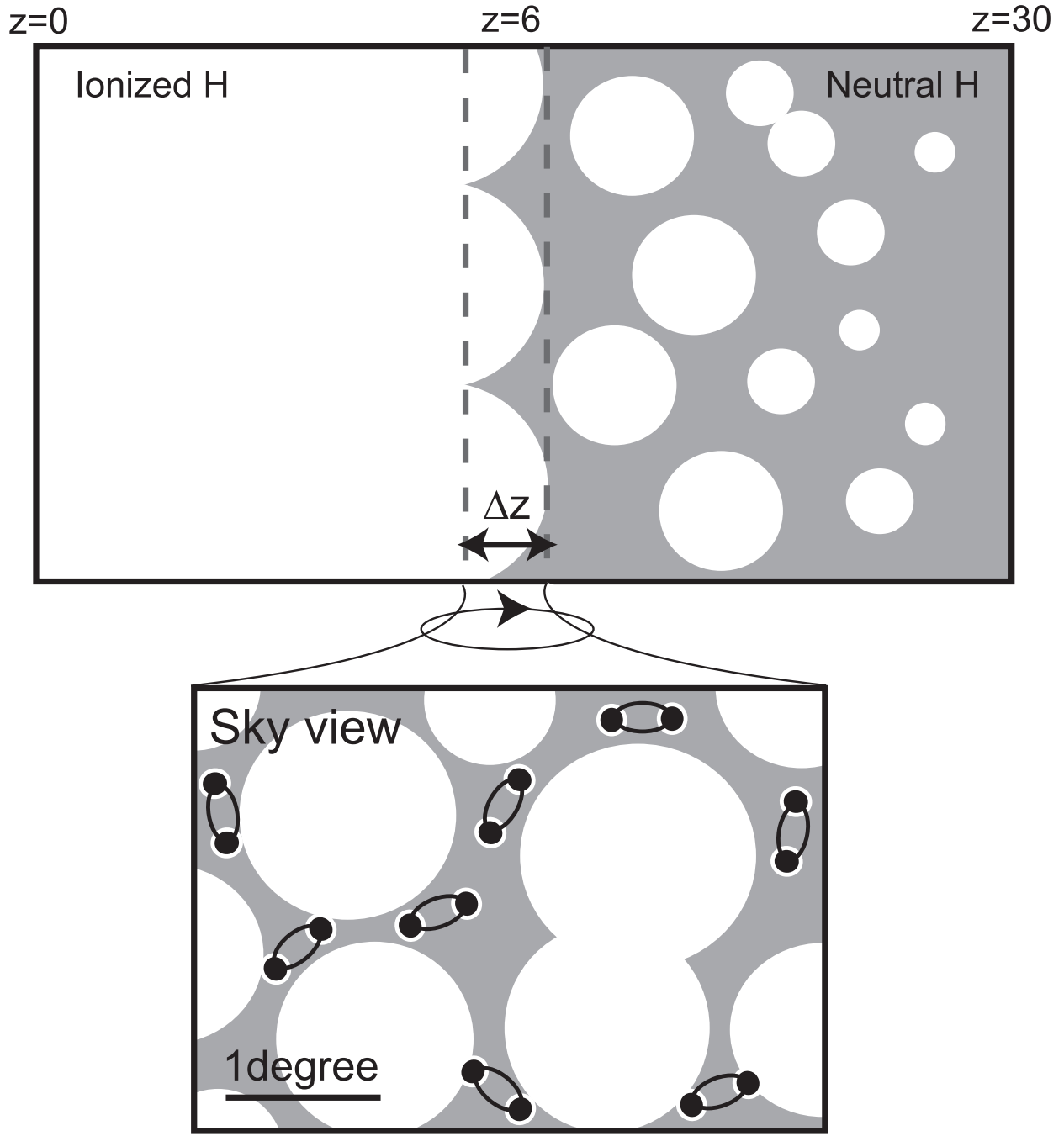}
\caption{\label{fig:reion}
Upper panel -- Schematic figure of reionized bubbles, with 
neutral regions shown as shaded zones.
Lower panel -- Schematic sky view at the end of the bubble overlap era.
Few MBH mergers take place inside an ionized bubble if the MBH binaries 
lose their angular momentum via gas drag, as the photoionizing background 
reduces the amount of MBH-dragging gas.
}
\end{figure}

Comparing equation (\ref{eq:finite}) with equation (\ref{eq:variance}),
we find that the scatter in the reionization redshift is
$\Delta z \sim 0.14$, and the ionized bubble has a comoving size
$R \sim 60$ Mpc ($\theta_{b} \sim 0.4^{\circ}$ on the sky)
at the end of reionization at $z\sim 6$.
This scale is much larger than the correlation length of host halos $r_{0}$,
and may be detectable through GWs.
Let us consider here a redshift shell $z \sim z+\Delta z$
at the end of reionization (see Figure~\ref{fig:reion}).
Since few mergers take place inside an ionized bubble
and the bubble separation is about their size $R\sim 60$ Mpc
($\theta_{b} \sim 0.4^{\circ}$ on the sky),
the MBH mergers should have a correlation $\xi \sim 1$ on this scale.
The number of merger pairs whose distances are less than $R$ 
(less than $\theta_{b}$ on the sky)
is approximately
\beqa
N_{\rm pair} \sim 
\left[\frac{dN_{\rm merg}}{dz dt_{\rm obs}} T_{\rm obs}
\Delta z \right]^{2}
\frac{\theta_{b}^{2}}{4}
\sim 20
\left(\frac{dN_{\rm merg}/dz dt_{\rm obs}}{10^{3}/{\rm yr}/z}\right)^{2}
\left(\frac{T_{\rm obs}}{10{\rm yr}}\right)^{2}
\left(\frac{\Delta z}{0.14}\right)^{2}
\left(\frac{\theta_{b}}{0.4^{\circ}}\right)^{2}
\eeqa
for an observation time $T_{\rm obs}$.
Since the Poisson error of the correlation function is 
$\Delta \xi/\xi \sim N_{\rm pair}^{-1/2}$,
the spatial correlation of MBH mergers 
may be detectable by GWs at a 4$\sigma$ level 
in the case (b) of Figure~\ref{fig:rate}.
(Even if the actual bubble size is less than $\sim 60$ Mpc
\citep{miralda00,gnedin00},
we may cross-correlate the merger position with the bubble position
to detect a correlation as discussed in \S~\ref{sec:diss}.)
To locate MBH mergers within an angle $\theta_{b}$
LISA may need EM counterparts to the MBH mergers,
while DECIGO/BBO, with a resolution better than $\theta_{b}$ 
\citep{takahashi03}, may do so without counterparts.

The bubble overlap may also be accompanied by a sharp drop in the merger 
rate because the photoionization effectively raises the minimum halo
mass that harbors a central MBH (e.g., from the case (b) to the case (a)
in Figure~\ref{fig:rate}).
Therefore the merger history may also be used to determine the
reionization redshift.\footnote{
\citet{wyithe03} also suggest that the merger rate drops after reionization.
However their reason is not the suppression of angular momentum losses
but the suppression of MBH formation.
}

\section{Discussion}\label{sec:diss}

A precise estimate of the reduction of the merger rate at reionization is 
beyond the scope of the paper, since it would depend on the MBH-dragging 
mechanism \citep{escala04} and the degree of self-shielding of ionizing 
photons \citep{susa04}. We note that gas will eventually evaporate even
if the gas is self-shielding in mini halos with $T_{\rm vir} \siml 10^4$
K. This is because the outside layer of gas is always ionized under the
ionizing background and can escape from halos.
We also note that the photoionization effect may be weak at high
redshift $z \simg 10$ \citep{dijkstra04}.

In the future, the spatial distribution of ionized 
bubbles will be determined by other methods such as 21 cm tomography 
\citep{tozzi00,ioka05}, Ly$\alpha$ spectroscopy \citep{miralda98} 
and dispersion measures \citep{ioka03}. For such measurements, the
cross-correlation of the merger position with the bubble position
(like the cross-correlation with LSSs outlined in \S~\ref{sec:align})
would make it easy to detect a signal. For example, if the fraction of ionized 
bubbles on the sky is $f_{b} \sim 1/3$ in a redshift shell $z \sim z+\Delta z$,
we only need $\sim 5^2 f_{b}^{-1} \sim 80$ events in $z \sim z+\Delta z$
to detect a 5$\sigma$ correlation. The rate of MBH mergers and the mechanism 
by which the angular momentum loss occurs may be independently constrained 
or determined through purely GW measurements, such as outlined in \S \ref{sec:gas}

In summary, we
have discussed two possible ways to exploit gravitational wave observations
in the $10^{-4}-10$ Hz range for gleaning information about the high redshift
LSS formation and about the mechanism of angular momentum loss in MBH mergers, 
as well as the reionization of the universe. The former would be indicated
by an alignment of the GW polarization direction with a principal axis
of the LSS, due to a preferential vector along which the galaxies approach.
The latter would be detectable through an MBH correlation length larger
than that of the host LSS, which can occur if the MBH angular momentum loss
occur through gas drag. In both cases we would gain information about the LSS
additional to and independent of any obtained from electromagnetic signals,
and in the latter also about the astrophysics of MBH mergers as well as an
independent measure of the reionization epoch.

\acknowledgments
We thank M.~J.~Rees for useful comments.  This work was supported in 
part by the Eberly Research Funds of Penn State and by the Center for 
Gravitational Wave Physics under grant PHY-01-14375 (KI), 
and NASA NAG5-13286, NSF AST 0307376 (PM).

%
%

%
%

\newpage

\newpage

\newpage

\end{document}